\documentclass[12pt]{article}

\usepackage{amsfonts,amsmath,amssymb,amsthm}
\usepackage{scalerel}[2016/12/29]
\usepackage{epsfig,graphicx}
\usepackage{tikz}
\usepackage{float}
\usepackage{braket}
\usepackage{url,hyperref}
\usetikzlibrary{decorations,arrows}
\usetikzlibrary{decorations.markings}
\usetikzlibrary{shapes.geometric}

\newcommand{\rotraise}[1]{%
  \StrLen{#1}[\slen]
  \forloop[-1]{idx}{\slen}{\value{idx}>0}{%
    \StrChar{#1}{\value{idx}}[\crtLetter]%
    \IfSubStr{tlQWERTZUIOPLKJHGFDSAYXCVBNM}{\crtLetter}
      {\raisebox{\depth}{\rotatebox{180}{\crtLetter}}}
      {\raisebox{1ex}{\rotatebox{180}{\crtLetter}}}}%
}

\def\bb{\hskip -0.5mm}
\def\be{\begin{equation}}
\def\ee{\end{equation}}
\def\bea{\begin{eqnarray}}
\def\eea{\end{eqnarray}}
\def\tr{{\rm tr}}

\def\rme{\mathrm{e}}
\def\rmi{\mathrm{i}}

\def\Q{\mathrm{Q}}
\def\rme{\mathrm{e}}
\def\rmi{\mathrm{i}}

\begin{document}

\thispagestyle{plain}

\title{\bf\Large Algebraic area enumeration of random walks on the  honeycomb lattice}

\author{Li Gan$^*$~{\scaleobj{0.9},} St\'ephane Ouvry$^*$~ {\scaleobj{0.9}{\rm and}} \ Alexios P. Polychronakos$^\dagger$}

\date{\today}

\maketitle

\begin{abstract}
We study the enumeration of closed walks of given length and algebraic area on the honeycomb lattice. Using an irreducible
operator realization of honeycomb lattice moves, we map the problem to a Hofstadter-like Hamiltonian and show that the
generating function of closed walks maps to the grand partition function of a system of particles with exclusion statistics
of order $g=2$ and an appropriate spectrum, along the lines of a connection previously established by two of the authors. Reinterpreting the results in terms
of the standard Hofstadter spectrum calls for a mixture of $g=1$ (fermion) and $g=2$ exclusion whose physical meaning and properties
require further elucidation. In this context we also obtain some unexpected Fibonacci sequences within the weights of
the combinatorial factors appearing in the counting of walks.
\end{abstract}

\noindent
* LPTMS, CNRS,  Universit\'e Paris-Sud, Universit\'e Paris-Saclay,\\ \indent 91405 Orsay Cedex, France; {\it li.gan92@gmail.com, stephane.ouvry@u-psud.fr}

\noindent
$\dagger$ Physics Department, the City College of New York, NY 10031, USA and \\ \indent
The Graduate Center of CUNY, New York, NY 10016, USA;
\\ \indent
{\it apolychronakos@ccny.cuny.edu}
\vskip 1cm

\vfill
\eject

\tableofcontents

\newpage

\section{Introduction}

{The algebraic area enumeration of closed random walks on two-dimensional lattices is a topic with rich mathematical and
physical implications since it has an intimate connection to discrete quantum models. The algebraic area is defined as the total
oriented area spanned by the walk as it traces the lattice. A unit lattice cell enclosed in a counter-clockwise (positive) direction
has an area $+1$, whereas when enclosed in a clockwise (negative) direction it has an area $-1$. The total algebraic area is the area enclosed by the walk weighted by its winding number: if the walk winds around more than once, the area is counted with multiplicity. Figure \ref{fig rw} represents examples of closed random walks on the square, triangular and honeycomb lattices. 
{\begin{figure}[H]
\centering 
{
\begin{tikzpicture}[scale=.7]
\draw[ultra thin, gray] (-2.6,-2.6) grid (2.6,2.6);
\draw[ultra thick,-=0.5,fill=green,fill opacity=0.5](-1,0)--(-1,1)--(-2,1)--(-2,0)--(-1,0);
\draw[ultra thick,-=0.5,fill=green,fill opacity=0.5](1,0)--(1,2)--(2,2)--(2,0)--(-2,0)--(1,0);
\draw[ultra thick,-=0.5,fill=red,fill opacity=0.5](-1,0)--(-1,-2)--(2,-2)--(2,-1)--(1,-1)--(1,0)--(-1,0);
\fill (0,0) circle (5pt);
\draw[black,ultra thick,-stealth](0,0)
{--++(0.7,0)};
\draw[black,ultra thick,-stealth](2,0)
{--++(0,1.2)};
{--++(0,-1.2)};
\draw[black,thick,-stealth](-0.8,-0.5)
{--++(0,1)};
\draw[black,thick,->=0.5](0.5,0.2)
{--++(1,0)};
\draw[black,ultra thick,-stealth](2,-2)
{--++(-1.7,0)};
\draw[black,ultra thick,-stealth](-1,1)
{--++(-0.7,0)};
{--++(0.7,0)};
\end{tikzpicture}
}
{
\begin{tikzpicture}[scale=.7]
\foreach \j in {0,1,...,2} {
\foreach \i in {0,1,...,4} {\draw[ultra thin, gray] (\i,1.732*\j)
{--++(60:1)--++(-60:1)--++(-1,0)--++(-60:1)--++(60:1)};
}}
\draw[ultra thin, gray](0,0.866){--++(5,0)};
\draw[ultra thin, gray](0,2.598){--++(5,0)};
\draw[black,thick,->=0.5](3.35,3.3)
{--++(-60:0.5)--++(0.5,0)};
\draw[black,thick,->=0.5](2.85,1.9)
{--++(60:0.5)--++(-0.5,0)};
\draw[black,ultra thick,-=0.5,fill=red,fill opacity=0.5](2,3.4641)
{--++(1,0) 
--++(-60:1) --++(1,0)
 --++(-120:1) --++(-60:1) --++(-120:1) --++(120:1) --++(-120:1) --++(120:1) --++(60:1) --++(60:1) --++(-1,0) --++(-120:1) --++(-120:1) 
 --++(120:1) --++(60:1) --++(60:1)};
\fill (2,3.4641) circle (5pt);
\draw[black,ultra thick,-=0.5,fill=green,fill opacity=0.5](1.5,0.859)
{--++(-120:1) --++(1,0) --++(120:1)};
\draw[black,ultra thick,-stealth](2,3.464)
{--++(0.8,0)};
\draw[black,ultra thick,-stealth](4,1.732)
{--++(-60:0.7)};
\draw[black,ultra thick,-stealth](2.5,0.866)
{--++(60:0.8)};
\draw[black,ultra thick,-stealth](2.5,2.598)
{--++(-120:1)};
\draw[black,ultra thick,-stealth](1,0)
{--++(0.7,0)};
\draw[black,ultra thick,-stealth](1,1.732)
{--++(60:1)};
\draw[black,thick,->=0.5](1.6,1.5)
{--++(-120:1)};
\end{tikzpicture}
}
{
\begin{tikzpicture}
\begin{scope}[scale=.42]
\clip (0,-0.05) rectangle (11,8.67);
\foreach \j in {0,1,...,4} {
\foreach \i in {0,1,...,3} {\draw[ultra thin, gray] (60:\j)++(120:\j)++(60:\i)++(-60:\i)++(\i,0)++(\i,0)
{--++(60:1)--+(120:1)++(0,0)--++(1,0)--++(60:1)--+(1,0)++(-120:1)--++(-60:1)};
}}
\draw[black,ultra thick,-=0.5,fill=green,fill opacity=0.5](60:1)++(120:1)++(60:1)++(-60:1)++(1,0)++(1,0)++(60:1)++(120:1)
{--++(-60:1) --++(1,0) --++(60:1) --++(1,0) --++(60:1) --++(1,0) --++(60:1) --++(1,0) --++(60:1) --++(120:1) --++(-1,0) --++(-120:1) --++(-1,0) --++(120:1) --++(-1,0) --++(-120:1) --++(-1,0) --++(-120:1) --++(-60:1) --++(-120:1)};
\fill (3,3.464) circle (8.33pt);
\draw[black,ultra thick,-stealth](6,3.464)
{--++(60:0.75)};
\draw[black,ultra thick,-stealth](6,6.928)
{--++(-0.8,0)};
\draw[ultra thin, gray,-](2,0)++(-0.0028,0){--++(1.0056,0)};
\draw[ultra thin, gray,-](5,0)++(-0.0028,0){--++(1.0056,0)};
\draw[ultra thin, gray,-](8,0)++(-0.0028,0){--++(1.0056,0)};
\end{scope} 
\end{tikzpicture}
}
\caption{Closed random walks of length ${\bf{n}}=20$ on the square, triangular and honeycomb lattice with algebraic area $-2$, $-12$ and $6$, respectively.}\label{fig rw}
\end{figure}
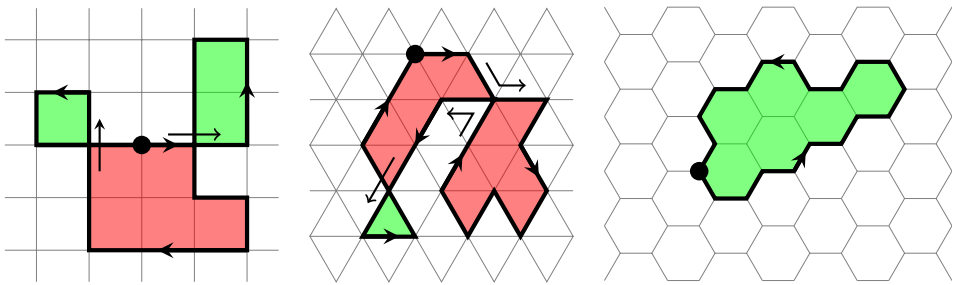}

In the case of the square lattice, the algebraic area enumeration is known to be embedded in the dynamics of the Hofstadter model \cite{Hofstadter} which describes the motion of an electron hopping on a square lattice in a uniform perpendicular magnetic field.  The generating function for the number $C_{\bf n}(A)$  of closed walks of length ${\bf n}=2n$ (necessarily even)  enclosing an algebraic area $A$ is given in terms of the trace of the Hofstadter Hamiltonian $H_{\gamma}$
\be\label{1}
\sum_A C_{{\bf{n}}}(A) \Q^{A } =  \text{ Tr} H_{\gamma}^{\bf n}
,\ee
where $\gamma = 2\pi\phi/\phi_0$ stands for the flux per plaquette in units of the flux quantum, $\Q={\rme}^{\rmi \gamma}$,
and $H_{\gamma}$  is the Hofstadter Hamiltonian
\be H_{\gamma}=u+u^{-1}+v+v^{-1}.\nonumber\ee
The unitary operators $u$ and $v$ are unit magnetic translations (hopping operators) in the $x$ and $y$ directions of the
square lattice and satisfy the ``quantum torus'' algebra
\be
v\; u = \Q\; u\; v
\label{qtorus}\ee
due to the perpendicular magnetic field piercing the lattice. Terms contributing to the trace  in (\ref{1}) must involve an equal number of $u$ and $u^{-1}$, and of
$v$ and $v^{-1}$. Such terms represent closed paths, each power of $H_{\gamma}$ representing one step. Because of the non commuting $u$ and $v$ in (\ref{qtorus}) the total
factor of $\Q$ for such paths can be seen to correspond to the algebraic area $A$ of the path, $v^{-1}u^{-1}vu = \Q$
corresponding to a path around an elementary plaquette. In quantum mechanics the trace becomes a sum of the expectation value of $H_\gamma$ over all quantum states, with an appropriate normalization.

In \cite{Ouvry and Wu} an explicit algebraic area enumeration was obtained in terms of a sum over compositions of the integer $n$. In \cite{Ouvry and Polychronakos 2019} and \cite{Ouvry and Polychronakos 2020}, an interpretation of this enumeration was given in terms of the statistical mechanics of particles obeying quantum \textit{exclusion statistics} with exclusion parameter $g$ ($g=0$ for bosons, $g=1$ for fermions, and higher $g$ means a stronger exclusion beyond Fermi). The square lattice enumeration was found to be governed by $g=2$ exclusion together with a Hofstadter-induced spectral function $s_k$ accounting for the 1-body quantum spectrum, whereas different types of lattice walks were governed by higher values of $g$
and, in general, other types of spectral functions. Explicit examples of such enumerations were  given, in particular for Kreweras-like chiral walks  on a triangular lattice \cite{Ouvry and Polychronakos 2019}, corresponding to yet another quantum Hofstadter-like model (chiral and non Hermitian, though) and $g=3$ exclusion. 
This particular chiral  model   is to be distinguished from the triangular lattice  Hofstadter-like model originally  proposed in \cite{Claro and Wannier}. Its butterfly structure  -- among other Hosftadter-like models --  has been studied in \cite{Yilmaz}.

An interesting case is the honeycomb lattice. It arises naturally in the form of graphene and carbon nanotubes, and many of
its quantum properties have  been extensively studied (see, for example, \cite{Novoselov, Bolotin, Perfetto}).
The honeycomb lattice is also relevant in graph theory \cite{Kang} and various physical models \cite{Kitaev,Wen,Herbut}. The quantum model for a particle hopping on the honeycomb lattice pierced by  a  perpendicular magnetic field was introduced in \cite{Rammal, Kreft and Seiler}. The effect of lattice defects on its spectrum was investigated in \cite{Pereira} and its  butterfly-like spectrum was obtained in \cite{Agazzi}.

In this work we address the question of the algebraic area enumeration of closed random walks on the honeycomb lattice:  can this enumeration be explicitly obtained, and does it fall in the category described in \cite{Ouvry and Polychronakos 2019} and \cite{Ouvry and Polychronakos 2020}, i.e., does it correspond to a particular  exclusion statistics? We will show that, indeed, the honeycomb enumeration can be interpreted in terms of $g=2$ exclusion provided that the Hofstadter
spectral function $s_k$ is ``diluted'' to a spectrum of alternating 1 and $s_k$. On the other hand, if we insist on using an
undiluted $s_k$, then $g=2$ exclusion has to be traded for a mixture of $g=1$ and $g=2$ exclusion whose physical meaning needs further clarification. In this process we will obtain some unexpected Fibonacci sequences, either for the number of compositions entering the enumeration or for the sum of the
coefficients weighting particular compositions, the occurrence of which remains to be better understood.
 
The paper is structured as follows: In Section \ref{Square lattice walks algebraic area enumeration} we review the Hofstadter model on the square lattice, where the coefficients of the secular determinant of the Hofstadter Hamiltonian  \cite{Kreft coefficient}  are  reinterpreted in terms of $g=2$ exclusion  partition functions. The algebraic area enumeration is then obtained in terms of the associated cluster coefficients. In Section \ref{Honeycomb lattice walks algebraic area enumeration} we study the honeycomb lattice and calculate the relevant partition functions and cluster coefficients, arriving at an explicit algebraic area enumeration expression. Some open questions are exposed in the Conclusions.}

\section{Square lattice walks algebraic area enumeration\label{Square lattice walks algebraic area enumeration}}
From now on we consider the flux $\gamma$ per lattice cell to be rational, i.e., 
$\phi/\phi_0=p/q$ with $p$ and $q$ co-prime, so $\Q=\exp(2 {\rmi} \pi p/q)$.

\subsection{Hofstadter Hamiltonian} 
\noindent When the magnetic flux is rational {the quantum torus algebra has a finite-dimensional irreducible
representation in which $u$ and $v$ are represented by the $q\times q$ ``clock'' and ``shift'' matrices \cite{Weyl}
\be u={\rme}^{{\rmi}k_y }\begin{pmatrix}
\Q & 0 & 0 & \cdots & 0 & 0 \\
0 & \Q^{2}  &0& \cdots & 0 & 0 \\
0 & 0 & \Q^{3}  & \cdots & 0 & 0 \\
\vdots & \vdots & \vdots & \ddots & \vdots & \vdots \\
0 & 0 & 0  & \cdots &\Q^{q-1}  & 0 \\
0 & 0 & 0  & \cdots & 0 & \Q^{q}\\
\end{pmatrix}, ~~
v = {\rme}^{{\rmi} k_x}\begin{pmatrix}
\;0\; & \;1\; & \;0\; & \;0\; & \cdots & \;0\; & \;0\; \\
0 & 0 & 1 & 0 &\cdots & 0 & 0 \\
0 & 0 & 0 & 1 &\cdots & 0 & 0\\
\vdots & \vdots & \vdots & \vdots &\ddots & \vdots & \vdots\\
0 & 0 & 0 & 0 &\cdots & 1 & 0\\
0 & 0 & 0 & 0 &\cdots &0& 1\\
1 & 0 & 0 & 0 & \cdots & 0& 0 \\
\end{pmatrix}
\label{uv}.\ee
$k_x\in[0,2\pi]$ and $k_y\in[0,2\pi]$ are the quasimomenta in the $x$ and $y$ lattice directions and are related to
the Casimirs of the $u,v$ algebra 
\be\nonumber
u^q={\rme}^{{\rmi} q k_y},~~v^q={\rme}^{{\rmi} q k_x}.
\ee
The Hofstadter Hamiltonian becomes the $q\times q $ matrix
\bea \nonumber H_q&=&\begin{pmatrix}
\Q {\rme}^{{\rmi} k_y}+\Q^{-1} {\rme}^{-{\rmi}k_y} & {\rme}^{{\rmi}k_x} & 0 & \cdots & 0 & {\rme}^{-{\rmi}k_x}\\
{\rme}^{-{\rmi}k_x} &\Q^2 {\rme}^{{\rmi}k_y}+\Q^{-2}{{\rme}^{-{\rmi}k_y}} & \quad \rme^{{\rmi}k_x}\quad & \cdots & 0 & 0 \\
0 & {\rme}^{-{\rmi}k_x} & () & \cdots & 0 & 0 \\
\vdots & \vdots & \vdots & \ddots & \vdots & \vdots \\
0 & 0 & 0 & \cdots & () & {\rme}^{{\rmi}k_x} \\
{\rme}^{{\rmi}k_x} & 0 & 0 & \cdots & \quad {\rme}^{-{\rmi}k_x}\quad & \Q^q {\rme}^{{\rmi}k_y}+\Q^{-q}{{\rme}^{-{\rmi}k_y} } \\
\end{pmatrix},
\nonumber\eea
\noindent {whose spectrum follows from the zeros of the secular determinant
$\det(1- z H_q)$, where $z $ denotes the inverse energy.
%

This secular determinant has been  shown \cite{Kreft coefficient} to  rewrite as
\bea \label{Hof det(1-zH)} \det(1- z H_q)=\sum_{n=0}^{\lfloor q/2 \rfloor} (-1)^nZ(n)z^{2n}-2\big(\cos(q k_x)+\cos(q k_y)\big)z^q,
\eea
where the $Z(n)$'s are given by the nested trigonometric sums} 
\bea
Z(n)=\sum_{k_1=0}^{q-2n}\sum_{k_2=0}^{k_1}\cdots\sum_{k_{n}=0}^{k_{n-1}}& 4\sin ^2\left(\displaystyle\frac{\pi (k_1+2n-1) p}{q}\right)4\sin ^2\left(\displaystyle\frac{\pi (k_2+2n-3) p}{q}\right)\cdots\nonumber \\ & 4\sin ^2\left(\displaystyle\frac{\pi  (k_{n-1}+3) p}{q}\right)4\sin ^2\left(\displaystyle\frac{\pi (k_{n}+1) p}{q}\right)
\label{Kreft}\eea
with $Z(0)=1$.

As we shall see, $Z(n)$  in (\ref{Kreft})  is at the core of the lattice walks algebraic area enumeration.
To recover (\ref{Kreft}) 
let us use an alternative form of the Hofstadter Hamiltonian involving a different but equivalent representation of the operators
$u$ and $v$, namely $-uv$ and $v$. They still satisfy the same quantum torus algebra  
\be\nonumber
v\; (-uv) = \Q\; (-uv)\; v,
\ee
albeit with a different Casimir $(-uv)^q = -{\rme}^{{\rmi} q (k_x + k_y )}$, and lead to the new Hamiltonian
\be H_q'=-uv-(uv)^{-1}+v+v^{-1},\nonumber\ee
i.e.,
{\footnotesize{\bea \nonumber & & H_q' =\begin{pmatrix}
 0  & (1-\Q^{}{\rme}^{{\rmi} k_y}){\rme}^{{\rmi} k_x} & 0 & \cdots & 0 & (1-\Q^{-q}{\rme}^{-{\rmi} k_y}){\rme}^{-{\rmi} k_x}\\
(1-\Q^{-1}{\rme}^{-{\rmi} k_y}){\rme}^{-{\rmi} k_x} & 0  &(1-\Q^{2}{\rme}^{{\rmi} k_y}){\rme}^{{\rmi} k_x} & \cdots & 0 & 0 \\
0 & (1-\Q^{-2}{\rme}^{-{\rmi} k_y}){\rme}^{-{\rmi} k_x} & 0  & \cdots & 0 & 0 \\
\vdots & \vdots & \vdots & \ddots & \vdots & \vdots \\
0 & 0 & 0  & \cdots & 0 & (1-\Q^{(q-1)}{\rme}^{{\rmi} k_y}){\rme}^{{\rmi} k_x} \\
(1-\Q^{q}{\rme}^{{\rmi} k_y}){\rme}^{{\rmi} k_x} & 0 & 0  & \cdots & (1-\Q^{-(q-1)}{\rme}^{-{\rmi} k_y}){\rme}^{-{\rmi} k_x} & 0 \\
\end{pmatrix},\nonumber \eea}}
or, denoting
$\omega(k)=(1-\Q^k {\rme}^{{\rmi}k_y}){\rme}^{{\rmi}k_x}$, 
\be 
H'_q = \begin{pmatrix}
0 & \omega(1) & 0 & \cdots & 0 & \bar{\omega}(q)\\
\bar{\omega}(1) & 0 &\omega(2) & \cdots & 0 & 0\\
0 & \bar{\omega}(2) & 0 & \cdots & 0 & 0\\
\vdots & \vdots & \vdots & \ddots & \vdots & \vdots\\
0 & 0 & 0 & \cdots & 0 & \omega(q-1) \\
\omega(q) & 0 & 0 & \cdots & \bar{\omega}(q-1) & 0 \\
\end{pmatrix}\nonumber\label{Hof Hq'}.
\ee
Its secular determinant is the same as that of $H_q$ given in (\ref{Hof det(1-zH)})  but for the new Casimirs, that is,
\bea \det(1-z H'_q)
&=& \sum_{n=0}^{\lfloor q/2 \rfloor}(-1)^nZ(n)z^{2n}-\left( \prod_{j=1}^{q}\omega(j)+\prod_{j=1}^{q}\bar{\omega}(j) \right)z^q\nonumber\\ 
&=& \sum_{n=0}^{\lfloor q/2 \rfloor}(-1)^nZ(n)z^{2n}-{2}\big(\cos(qk_x)-\cos(qk_x+qk_y)\big)z^q.
\label{Hof det(1-zH')} \eea
Let us  set $\omega(q)=0$, which makes the cosine term in (\ref{Hof det(1-zH')}) vanish and the matrix
$H'_q$ tridiagonal}
{\bea & & H_q'|_{\omega(q)=0} =\begin{pmatrix}
0  & (1\bb-\bb \Q^{1-q}){\rme}^{{\rmi}k_x} & 0 & \cdots & 0 & 0\\
(1\bb -\bb \Q^{q-1}){\rme}^{-{\rmi}k_x} & 0  &(1\bb -\bb \Q^{2-q}){\rme}^{{\rmi}k_x} & \cdots & 0 & 0 \\
0 & (1\bb-\bb \Q^{q-2}){\rme}^{-{\rmi}k_x} & 0  & \cdots & 0 & 0 \\
\vdots & \vdots & \vdots & \ddots & \vdots & \vdots \\
0 & 0 & 0  & \cdots & 0 & (1\bb -\bb \Q^{-1}){\rme}^{{\rmi}k_x}\\
0 & 0 & 0  & \cdots & (1\bb -\bb \Q^{}){\rme}^{-{\rmi}k_x} & 0 \\
\end{pmatrix}.\nonumber\\\label{Hof Hq'bis}\nonumber\eea
This form provides an iterative procedure for calculating the $Z(n)$'s. Putting aside for a moment that
$\Q=\exp(2{\rmi}\pi p/q)$ and leaving it as a free parameter, we introduce the spectral function
\be
s_{k}= (1-\Q^{k})(1-\Q^{-k}).
\label{sk}\ee
Denoting the secular determinant $\det(1-z H'_q|_{\omega(q)=0})=d_q$, its expansion
in terms of the first row yields
\be 
d_q=d_{q-1}-z^2 s_{q-1} \, d_{q-2},~~~ q\ge 2,
\label{Hofrec}\ee
{where, by convention, $d_0=d_1=1$}. Expanding $d_q$ as a polynomial in $z$ and solving the corresponding recursion relation for its coefficients,
we obtain (see Appendix A)
\bea
Z(n) 
 = \sum_{k_1=1}^{q-2n{+1}} \sum_{k_2=1}^{k_1} \cdots \sum_{k_{n}=1}^{k_{n-1}}
s_{k_1+2n-2}
s_{k_2+2n-4} \cdots s_{k_{n-1}+2}
s_{k_{n}}\label{Hof Zn bis},
\eea
which, upon restoring $\Q$ to its actual  value $\exp(2{\rmi}\pi p/q)$, i.e., the spectral function $s_k$ to its actual form $s_k= 4 \sin^2 ({\pi k p / q})$, gives (\ref{Kreft}).

The recursion {(\ref{Hofrec})} is at the root of the connection between square lattice walks
and $g=2$ exclusion statistics. Interpreting the spectral function $s_k$ as the Boltzmann factor for a 1-body
level $\rme^{-\beta \epsilon_k}$ and $-z^2$ as the fugacity  $z'$, 
{(\ref{Hofrec})} {can be interpreted as an expansion of a grand partition function ${\cal Z}_{q-1}$ -- here identified with $d_q$ --
of noninteracting particles in  $q-1$  quantum levels $\epsilon_1 , \dots , \epsilon_{q-1}$, obeying the exclusion principle  that no two particles can occupy adjacent levels, namely
 \be {\cal Z}_{q-1}={\cal Z}_{q-2}+z' s_{q-1} \, {\cal Z}_{q-3}\nonumber \ee
in terms of the last level $\epsilon_{q-1}$ being empty (first term) or
occupied (second term)}. Then (\ref{Hof det(1-zH')}) identifies $Z(n)$ as the $n$-body partition function for particles occupying these $q-1$ quantum states, with  gaps of 2 between successive terms reproducing $g=2$ exclusion.

\subsection{Algebraic area enumeration on the square lattice \label{section enumeration on square lattice}}

{As already stressed, when  $\Q = \exp(2{\rmi}\pi p/q)$  the algebraic area counting (\ref{1}) }
\be\label{Hof Cn and Tr}
\sum_A C_{{\bf{n}}}(A) \Q^{A } = {1\over q} \text{Tr} H_{q}^{\bf n} 
\ee
involves a trace over a finite number $q$ of quantum states.
To normalize the contribution of each
path to $\Q^A$ and reproduce the left-hand side of (\ref{Hof Cn and Tr}), a factor of $1/q$  must be included in the normalization.
Also, when ${\bf n} \ge q$ the trace involves extra  terms arising from the Casimirs $k_x,k_y$
similar to the
cosine terms in (\ref{Hof det(1-zH)}), corresponding to open paths that close only up to periods $(q,q)$ on the lattice
(``umklapp'' terms on the quantum torus). These spurious contributions can be eliminated by integrating the Casimirs
$k_x$ and $k_y$ over  $[0,2\pi]$ which makes all factors of ${\rme}^{{\rmi}qk_x}$ and $\rme^{{\rmi}qk_y}$
vanish. So the definition of the trace in (\ref{Hof Cn and Tr}) is
\be
\text{Tr} H_q^{\bf n}=\frac{1}{(2\pi)^2}\int_{0}^{2\pi} \bb\bb\bb d k_x \int_{0}^{2\pi} \bb\bb\bb dk_y \;
\text{tr} H_q^{\bf n}
,\label{fulltrace}\nonumber\ee
which corresponds to summing over the $q$ bands of the spectrum and  over the
scattering states labeled by $k_x,k_y$, in a continuum normalization. 

To relate this trace to the $Z(n)$'s in (\ref{Kreft}) or, equivalently, in (\ref{Hof Zn bis}), one has to use
\be\label{trlog}\nonumber
\log \det(1 -z H_q) =\tr \log(1 -z H_q) = -\sum_{n=1}^{\infty} {z^n \over n} \,\tr H_q^{n},
\ee
and the fact that, in statistical mechanics, the  $Z(n)$ are viewed as $n$-body partition functions  with cluster coefficients $b(n)$ defined via the grand partition function $\sum_{n=0}^{\infty}Z(n)z^n$}
\be\log\left(\sum_{n=0}^{\infty}Z(n) z^n \right)=\sum_{n=1}^{\infty} b(n) z^n\label{00}
\ee
 with $z$ playing the role of the fugacity.
Trading $z$  for $-z^2$ in (\ref{00}), keeping in mind that trivially $\tr H_q^{2n+1}=0$, and putting everything together we reach the conclusion \cite{Ouvry and Wu, 
Ouvry and Polychronakos 2019} that the trace in (\ref{Hof Cn and Tr}) for ${\bf n}=2n$ is nothing but the cluster coefficient $b(n)$ up to a trivial factor
\be \label{i}\text{Tr} H_q^{\bf n}=2n(-1)^{n+1} b(n).\ee
{\noindent The cluster coefficients can in turn be directly read from the $Z(n)$'s in (\ref{Hof Zn bis}): one gets
\iffalse. The first few $b(n)$'s are}
\begin{align} \nonumber 
b(1)&=\sum_{k=1}^q s_{k}\;,\\\nonumber
 -b(2)&=\frac{1}{2}\sum_{k=1}^q s_{k}^2+\sum_{k=1}^{q-1} s_{k+1}s_{k}\;,\\
 \nonumber b(3)&=\frac{1}{3}\sum_{k=1}^q {s}_{k}^3+\sum_{k=1}^{q-1} {s}_{k+1}^2{s}_{k}+\sum_{k=1}^{q-1} {s}_{k+1}{s}_{k}^2+\sum_{k=1}^{q-2} {s}_{k+2}{s}_{k+1}{s}_{k}\;,\\
 \nonumber -b(4)&=  \frac{1}{4} \sum_{k=1}^q {s}_{k}^4+ \sum_{k=1}^{q-1} {s}_{k+1}^3 {s}_{k}+ \sum_{k=1}^{q-1} {s}_{k+1} {s}_{k}^3 +\frac{3}{2}  \sum_{k=1}^{q-1} {s}_{k+1}^2 {s}_{k}^2\\\nonumber & + 2  \sum _{k=1}^{q-2} {s}_{k+2} {s}_{k+1}^2 {s}_{k}+ \sum_{k=1}^{q-2} {s}_{k+2}^2
   {s}_{k+1} {s}_{k} + \sum_{k=1}^{q-2} {s}_{k+2} {s}_{k+1}
   {s}_{k}^2\\& + \sum_{k=1}^{q-3} {s}_{k+3} {s}_{k+2} {s}_{k+1} {s}_{k}\nonumber\;,
\end{align}
{which can be generalized as}\fi
\be 
b(n)=(-1)^{n+1}\hskip -0.3cm \sum_{\substack{l_1, l_2, \ldots, l_{j} \\ { \rm composition}\;{\rm of}\;n}} \hskip -0.4cm 
c(l_1,l_2,\ldots,l_{j} )\sum _{k=1}^{{q-j}} s^{l_{j}}_{k+j-1}\cdots s^{l_2}_{k+1} {s}^{l_1}_k,\label{Hof bn}
\ee 
{where the $c(l_1,l_2,\ldots,l_{j})$'s are labeled by the compositions of the integer  ${{n}}$ with
\be c(l_1,l_2,\ldots,l_{j})= \frac{{l_1+l_2\choose l_1}}{l_1+l_2}\;\; l_2\frac{{l_2+l_3\choose l_2}}{l_2+l_3}\;\cdots \;\; l_{{j}-1}\frac{{l_{{j}-1}+l_{j}\choose l_{{j}-1}}}{l_{{j}-1}+l_{j}}.\label{Hof coef c}\ee 
Further, the trigonometric sums $ \frac{1}{q}\sum _{k=1}^{{q-j}} s^{l_{j}}_{k+j-1}\cdots{{s}^{l_2}_{k+1}}{s}^{l_1}_k$ can also be computed  \cite{Ouvry and Wu, Ouvry and Polychronakos 2020}
{\footnotesize
\bea \nonumber {1\over q}\sum _{k=1}^{{q-j}} s^{l_{j}}_{k+j-1}\cdots{{s}^{l_2}_{k+1}}{s}^{l_1}_k=&&{ \sum_{A=-\infty}^{+\infty}\cos\left(\frac{2A\pi p}{q}\right)}\\
&&\hskip -4.1cm\sum_{k_3=-l_3}^{l_3}\sum_{k_4=-l_4}^{l_4}\cdots\sum_{k_{j}=-l_j}^{l_{j}}{2l_1\choose {l_1+A+\sum_{i=3}^{j}(i-2)k_i}}{2l_2\choose {l_2-A-\sum_{i=3}^{j}(i-1)k_i}}\prod_{i=3}^{j}{2l_i\choose l_i+k_i}.\label{trig}\eea}
Using (\ref{i}), (\ref{Hof bn}), (\ref{Hof coef c}) and (\ref{trig}) and  keeping in mind that ${\bf n} = 2n$, we deduce the desired algebraic area counting
\be \nonumber \sum_A C_{\bf n}(A)\Q^A=\frac{1}{q} \text{ Tr} H_q^{\bf n}=2n \sum_{\substack{l_1, l_2, \ldots, l_{j}\\ { \rm composition}\;{\rm of}\;n}} \hskip -0.4cm 
c(l_1,l_2,\ldots,l_{j} )\frac{1}{q}\sum _{k=1}^{{q-j}} s^{l_{j}}_{k+j-1}\cdots s^{l_2}_{k+1} {s}^{l_1}_k,\ee
i.e.,
{\small{
\bea \nonumber
C_{\bf n}(A)=&& 2n\hskip -0.3cm \sum_{\substack{l_1, l_2, \ldots, l_{j} \\ { \rm composition}\;{\rm of}\;n}} \hskip -0.4cm 
c(l_1,l_2,\ldots,l_{j})\\ \nonumber
&& \sum_{k_3=-l_3}^{l_3}\sum_{k_4=-l_4}^{l_4}\cdots\sum_{k_{j}=-l_j}^{l_{j}}{2l_1\choose {l_1+A+\sum_{i=3}^{j}(i-2)k_i}}{2l_2\choose {l_2-A-\sum_{i=3}^{j}(i-1)k_i}}\prod_{i=3}^{j}{2l_i\choose l_i+k_i}.\\
\label{countingcount}\eea}
}We also note that, since
\be \sum_{\substack{l_1, l_2, \ldots, l_{j}\\ {\rm composition}\;{\rm of}\;n}} \hskip -0.3cm c(l_1,l_2,\ldots,l_{j})=\frac{{2n\choose n}}{{2n}},\nonumber \ee
and, when $q\to\infty$ { \cite{Ouvry and Wu, Ouvry and Polychronakos 2019}},
\be {\frac{1}{q}\sum _{k=1}^{q-j} s^{l_{j}}_{k+j-1}\cdots s^{l_2}_{k+1} {s}^{l_1}_k} \to {2(l_1+l_2+\ldots+l_j)\choose l_1+l_2+\ldots+l_j}\label{counting},\ee
the overall closed square lattice walks counting \be 2n \sum_{\substack{l_1, l_2, \ldots, l_{j}\\ { \rm composition}\;{\rm of}\;n}} \hskip -0.4cm 
c(l_1,l_2,\ldots,l_{j} ) {2(l_1+l_2+\ldots+l_j)\choose l_1+l_2+\ldots+l_j}={{{2n}}\choose {{{n}}}}^2={{\bf{n}}\choose {{\bf{n}}/2}}^2\nonumber\ee
is recovered as it should (see  Appendix {B} for some enumeration  examples).

\section{Honeycomb lattice walks algebraic area enumeration\label{Honeycomb lattice walks algebraic area enumeration}}
We plan to follow the same route as above to obtain an explicit algebraic area enumeration for closed walks on the honeycomb lattice.

\subsection{Honeycomb Hamiltonian\label{Honeycomb Hamiltonian}}

{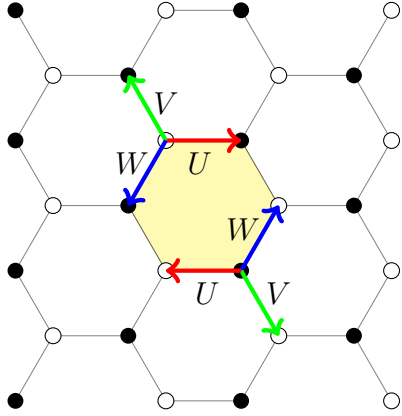
\begin{figure}[H]
\centering
{
\begin{tikzpicture}
\begin{scope}[scale=1]
\clip (-0.11,-0.11) rectangle (5.11,5.31);
\foreach \j in {0,1,...,2} {
\foreach \i in {0,1,...,1} {\draw[ultra thin, gray] (60:\j)++(120:\j)++(60:\i)++(-60:\i)++(\i,0)++(\i,0)
{--++(60:1)--+(120:1)++(0,0)--++(1,0)--++(60:1)--+(1,0)++(-120:1)--++(-60:1)};
}}
\draw [ultra thin,transparent,fill=yellow,opacity=0.3] (2,1.73205)--(3,1.73205)--(3.5,2.59808)--(3,3.46410)--(2,3.46410)--(1.5,2.59808)--(2,1.73205);
\draw[ultra thin, gray] (2,0) {--++(1,0)};
\foreach \j in {0,1,...,3} {
\draw[fill=white] (0.5,-0.86602)++(0,\j*1.73205) circle (3pt);
\draw[fill=white] (2,0)++(0,\j*1.73205) circle (3pt);
\draw[fill=white] (3.5,-0.86602)++(0,\j*1.73205) circle (3pt);
\draw[fill=white] (5,0)++(0,\j*1.73205) circle (3pt);
\fill[fill=black] (0,\j*1.73205) circle (3pt);
\fill[fill=black] (1.5,-0.86602)++(0,\j*1.73205) circle (3pt);
\fill[fill=black] (3,\j*1.73205) circle (3pt);
\fill[fill=black] (4.5,-0.86602)++(0,\j*1.73205) circle (3pt);
}
\tikzset{
  big arrow/.style={
    decoration={markings,mark=at position 1 with {\arrow[scale=1.3,#1]{>}}},
    postaction={decorate},
    shorten >=1pt}}
\draw[blue,ultra thick,big arrow](60:1)++(1,0)++(60:1)++(1,0)node[black] at ++(-0.45,-0.3) {$U$}{--++(60:1)};
\draw[green,ultra thick,big arrow](60:1)++(1,0)++(60:1)++(1,0)node[black] at ++(0.5,-0.3) {$V$}{--++(-60:1)};
\draw[red,ultra thick,big arrow](60:1)++(1,0)++(60:1)++(1,0)node[black] at ++(0.03,0.55) {$W$}{--++(-1,0)};
\draw[red,ultra thick,big arrow](60:1)++(1,0)++(60:1)++(120:1)++(60:1)node[black] at ++(0.45,-0.3) {$U$}{--++(1,0)};
\draw[green,ultra thick,big arrow](60:1)++(1,0)++(60:1)++(120:1)++(60:1)node[black] at ++(0,0.5) {$V$}{--++(120:1)};
\draw[blue,ultra thick,big arrow](60:1)++(1,0)++(60:1)++(120:1)++(60:1)node[black] at ++(-0.45,-0.3) {$W$}{--++(-120:1)};
\end{scope} 
\end{tikzpicture}
}
\caption{Hopping operators $U$, $V$, $W$  on the honeycomb lattice with $U^2=V^2=W^2=1$ and $(UVW)^2=\Q$.}\label{fig honeycomb lattice walks}
\end{figure}}

Consider a particle hopping on a honeycomb lattice pierced by a constant magnetic field  ({see Fig.~\ref{fig honeycomb lattice walks}}). The lattice  is bipartite with unitary
operators $U,V,W$ generating the  hoppings in each direction   and such that when the particle hops around a
honeycomb cell it picks up a phase $\Q$ due to the magnetic field. They satisfy the honeycomb algebra
\be 
U^2=V^2=W^2=1,~~ (UVW)^2=\Q.
\label{honey}\ee
The Hofstadter-like Hamiltonian follows as} 
\bea \nonumber H_{\text{honeycomb}} = a U+ b V+ c W,\eea
with $a,b,c \in {\mathbb{R}}^+$ transition amplitudes. The physical Hilbert space consists of the irreducible representations
of the honeycomb algebra. As in the square lattice case, the quasimomenta are encoded in the Casimirs of the algebra.

In the case of  an isotropic lattice, $a=b=c=1$, and a rational flux, $\Q=\exp(2 {\rmi} \pi p/q)$ with $p$ and $q$ co-prime, the irreducible representation of $U$, $V$ and $W$ for generic quasimomenta
(Casimirs) becomes  $2q$-dimensional (see Appendix C)
\bea \nonumber U = \begin{pmatrix}
\hskip -0.25cm 0 & u \\
u^{-1} & 0 
\end{pmatrix}
,~~
V = \begin{pmatrix}
\hskip -0.25cm0 & v\\
v^{-1} & 0
\end{pmatrix}
,~~ W = \begin{pmatrix}
\hskip -0.25cm 0 & \Q^{1/2} v u^{-1} \\
\Q^{-1/2} u v^{-1}  & 0
\end{pmatrix}\nonumber
\eea
with $u,v$ given in (\ref{uv}), and the honeycomb Hamiltonian reduces to the $2q\times 2q$ matrix
\bea  H_{2q} = 
\begin{pmatrix}
\hskip -0.25cm 0 & u\bb+\bb v\bb +\bb\Q^{1/2} v u^{-1} \\
u^{-1}\bb+\bb v^{-1}\bb+\bb\Q^{-1/2} u v^{-1} & 0 
\end{pmatrix} = \begin{pmatrix}
0 & A \\
A^\dagger & 0 
\end{pmatrix}
\label{it}. \eea
\noindent
{Its square is block-diagonal}
\bea H_{2q}^2  = \begin{pmatrix}
A A^\dagger & 0 \\
0 & A^\dagger\bb A 
\end{pmatrix}=
\begin{pmatrix}
H_{q} & 0 \\
0 & \tilde{H}_{q} 
\end{pmatrix}
,\nonumber\eea
where ${H}_{q}= A A^\dagger$ and $\tilde{H}_{q} = A^\dagger\bb A$ have identical spectra 
equal to
the square of the honeycomb Hamiltonian spectrum. Denoting
\be \omega(k)=\Q^{-k}\big(1+{\rme}^{-{\rmi}k_y}\Q^{\frac{1}{2}-k}\big){\rme}^{-{\rmi}(k_y-k_x)}\nonumber,\ee
$H_q$   can be rewritten as
\bea \label{hex Hq}  H_q=\begin{pmatrix}
 1+\omega(2)\bar \omega(2) & \omega(2) & 0 & \cdots & 0 & \bar \omega(1)\\
\bar \omega(2) & 1+\omega(3)\bar \omega(3) & \quad \omega(3)\quad & \cdots & 0 & 0 \\
0 & \bar \omega(3) & () & \cdots & 0 & 0 \\
\vdots & \vdots & \vdots & \ddots & \vdots & \vdots \\
0 & 0 & 0 & \cdots & () & \omega(q)\\
\omega(1) & 0 & 0 & \cdots & \quad \bar \omega(q)\quad & 1+\omega(1)\bar \omega(1) \\
\end{pmatrix}
\eea
with secular determinant
\bea \nonumber
&& \det(1-z H_{2q})
=\det(1-z^2 H_q )\\ 
&&=\sum_{n=0}^{q} (-1)^nZ(n)z^{2n}+\bigg((-1)^q \prod_{j=1}^q \omega(j)\bar{\omega}(j)-\prod_{j=1}^q \omega(j)-\prod_{j=1}^q \bar{\omega}(j)\bigg)z^{2q}\nonumber\\ 
&&=\sum_{n=0}^{q} (-1)^nZ(n)z^{2n}+2\bigg(\hskip -0.1cm -\Q^{\frac{q}{2}}\big(\cos(q k_x -2q k_y)+ \cos(q k_y)\big) + (-1)^q\big(\cos(q k_x - q k_y) +1\big)\bigg)z^{2q}.\nonumber\\ \label{hex det}\eea

\subsection{Honeycomb coefficients $Z(n)$}
Our aim is to find  for the $Z(n)$ in (\ref{hex det}) an expression  analogous to the one in (\ref{Kreft}) or (\ref{Hof Zn bis}) obtained in the Hofstadter case. To this end, we
reduce the honeycomb matrix (\ref{hex Hq}) to a tridiagonal form by making both corners $\omega(1)$ and $\bar \omega(1)$ vanish, i.e., by setting ${\rme}^{-{\rmi}k_y}=-\Q^{\frac{1}{2}}$ so that
$\omega(k)$ becomes
\be
\omega(k)|_{\omega(1)=0}= -\Q^{\frac{1}{2}-k}\left(1-\Q^{1-k}\right){\rme}^{{\rmi}k_x}\nonumber,
\ee 
and 
{\footnotesize \bea \hskip -2cm\nonumber {H_q}\bigg|_{\omega(1)=0}&\bb\bb\bb=\begin{pmatrix}
1+(1\bb-\bb\Q^{-1})(1\bb-\bb\Q)  & -\Q^{-\frac{3}{2}}(1\bb-\bb\Q^{-1}){\rme}^{{\rmi} k_x} & 0 & \cdots & 0 & 0\\
-\Q^{\frac{3}{2}}(1\bb-\bb\Q){\rme}^{-{\rmi} k_x} & 1+(1\bb-\bb\Q^{-2})(1-\Q^{2})  &-\Q^{-\frac{5}{2}}(1\bb-\bb\Q^{-2}){\rme}^{{\rmi} k_x} & \cdots & 0 & 0 \\
0 &-\Q^{\frac{5}{2}}(1\bb-\bb\Q^2){\rme}^{-{\rmi} k_x} & ()  & \cdots & 0 & 0 \\
\vdots & \vdots & \vdots & \ddots & \vdots & \vdots \\
0 & 0 & 0  & \cdots & () & -\Q^{\frac{1}{2}-q}(1\bb-\bb\Q^{-(q-1)}){\rme}^{{\rmi} k_x} \\
0 & 0 & 0  & \cdots & -\Q^{q-\frac{1}{2}}(1\bb-\bb\Q^{q-1}){\rme}^{-{\rmi} k_x} &  1+(1\bb-\bb\Q^{-q})(1-\Q^{q}) \\
\end{pmatrix}.\\\label{hex Hq w(1)=0}\nonumber
\eea}
\hskip -0.2cm 
This also eliminates the $z^{2q}$ umklapp term in (\ref{hex det}), i.e., the secular determinant reduces to
\bea \nonumber
& &\det\left(1-z^2 H_q|_{\omega(1)=0}\right)
=\sum_{n=0}^{q} (-1)^nZ(n)z^{2n}.\eea
{Let us now consider $\Q$ as a free parameter and denote $d_q= \det(1-z^2 H_q|_{\omega(1)=0})$. Then expanding $d_q$ in terms of its bottom row we obtain the recursion relation
\be 
d_q=\left(1-\left[1+(1-\Q^{q})(1-\Q^{-q})\right]z^2\right)d_{q-1}-z^4(1-\Q^{q-1})(1-\Q^{-(q-1)})d_{q-2},~~q\ge 1,
\nonumber\ee
i.e.,
\be d_q=\Big(1-\big(1+s_{q}\big)z^2\Big)d_{q-1}-z^4 s_{q-1} d_{q-2},\label{hexrec}
\ee
with $d_0=1$, $d_j=0$ for $j<0$, and $s_k$ as in (\ref{sk}).
From (\ref{hexrec}) we can iteratively derive the $Z(n)$ {(see Appendix D)}. 

The above recursion  admits a simple $g=2$
exclusion statistics interpretation. Consider a set of $2q$ energy levels with spectral parameters $S_n$, $n=1,2,\dots,2q$ given by
\be\nonumber
S_{2k-1} = 1,~~~S_{2k} = s_k,
\ee
that is, $s_k$ ``diluted'' by unit insertions: $1,s_1,1,s_2,\dots,1,s_q$, and consider the grand partition function of $g=2$ exclusion
particles in the above spectrum $S_n$ with fugacity parameter $z$. Calling ${\cal Z}_{1,n}$ the truncated grand partition function
for levels $S_1,S_2,\dots,S_n$ and expanding it in terms of the last level $n$ being empty or filled, we obtain the recursion relations
\bea
n=2k :~~~~~~\,  &&{\cal Z}_{1,2k} = {\cal Z}_{1,2k-1} + z s_k {\cal Z}_{1,2k-2},\nonumber\\
n=2k\bb-\bb 1 :~~  &&{\cal Z}_{1,2k-1} = {\cal Z}_{1,2k-2} + z {\cal Z}_{1,2k-3}.\nonumber
\eea
From the $n=2k$ relation we can express the odd functions ${\cal Z}_{2k-1}$ in terms of even ones,
${\cal Z}_{1,2k-1} = {\cal Z}_{1,2k} - z s_k {\cal Z}_{1,2k-2}$.
Substituting this expression in the $n=2k-1$ relation and rearranging we obtain
\be
{\cal Z}_{1,2k} = (1+z+z s_k ) {\cal Z}_{1,2k-2} - z^2 s_{k-1} {\cal Z}_{1,2k-4}.
\nonumber\ee
This is identical to the recursion (\ref{hexrec}) upon shifting $z \to -z^2$ and identifying ${\cal Z}_{1,2k} = d_k$. Moreover, 
${\cal Z}_{2k}$ satisfies the same initial conditions as $d_k$, namely ${\cal Z}_{1,0} =1$, ${\cal Z}_{1,2k}=0$ for $k<0$. Therefore,
$d_q = {\cal Z}_{2q}$.

It follows that the expressions for the $n$-body partition functions $Z(n)$ and the cluster coefficients $b(n)$ are  identical to the corresponding expressions (\ref{Hof Zn bis})  and (\ref{Hof bn})  for 
square lattice  walks but now, instead of the spectrum $s_k$,  one has to consider the diluted spectrum $S_k$,  $k=1,\ldots,{ 2q}$ {(but note that $S_{2q} = s_q =0$, so the levels effectively end at $S_{2q-1} =1$)}
\bea
Z(n) 
 = \sum_{k_1=1}^{2q-2n{ +2}} \sum_{k_2=1}^{k_1} \cdots \sum_{k_{n}=1}^{k_{n-1}}
S_{k_1+2n-2}
S_{k_2+2n-4} \cdots S_{k_{n-1}+2}
S_{k_{n}},\label{honeycomb Zn diluted}\nonumber
\eea
\be 
b(n)=(-1)^{n+1}\hskip -0.3cm \sum_{\substack{l_1, l_2, \ldots, l_{j} \\ { \rm composition}\;{\rm of}\;n}} \hskip -0.4cm 
c(l_1,l_2,\ldots,l_{j} )\sum _{k=1}^{2q-j+1} S^{l_{j}}_{k+j-1}\cdots S^{l_2}_{k+1} {S}^{l_1}_k\label{honeycomb bn diluted}\nonumber
\ee
with  the same Hofstadter combinatorial factors $c(l_1,l_2,\dots,l_j )$ given in (\ref{Hof coef c}).
The corresponding diluted trigonometric sums ${1\over q}\sum _{k=1}^{{ 2q-j+1}} S^{l_{j}}_{k+j-1}\cdots{S}^{l_2}_{k+1}{S}^{l_1}_k $ can be expressed as
{\scriptsize
\bea \nonumber
&&{1\over q}\sum _{k=1}^{{ 2q-j+1}} S^{l_{j}}_{k+j-1}\cdots{S}^{l_2}_{k+1}{S}^{l_1}_k = \sum_{A=-\infty}^{+\infty}\cos\left( \frac{2A\pi p}{q}\right)
\\ \nonumber
&&\Bigg( \sum_{k_5=-l_5}^{l_5}\sum_{k_7=-l_7}^{l_7}\cdots  \sum_{k_{2\lfloor(j-1)/2\rfloor+1}=-l_{2\lfloor(j-1)/2\rfloor+1}}^{l_{2\lfloor(j-1)/2\rfloor+1}}{2l_1\choose {l_1+A+\displaystyle\sum_{\substack{i=5\\ i\text{ odd}}}^{2\lfloor(j-1)/2\rfloor+1}(i-3)k_i}/2}{2l_3\choose {l_3-A-\displaystyle\sum_{\substack{i=5\\ i\text{ odd}}}^{2\lfloor(j-1)/2\rfloor+1}(i-1)k_i/2}} \prod_{\substack{i=5\\ i\text{ odd}}}^{2\lfloor (j-1)/2\rfloor+1} {2l_i\choose l_i+k_i}
\\ \nonumber
&& + \sum_{k_6=-l_6}^{l_6}\sum_{k_8=-l_8}^{l_8}\cdots {\hskip -0.1cm} \sum_{k_{2\lfloor j/2\rfloor}=-l_{2\lfloor j/2\rfloor}}^{l_{2\lfloor j/2\rfloor}} {\hskip -0.1cm} {2l_2\choose {l_2+A+\displaystyle\sum_{\substack{i=6\\ i\text{ even}}}^{2\lfloor j/2\rfloor}(i-4)k_i/2}}{2l_4\choose {l_4-A-\displaystyle\sum_{\substack{i=6\\ i\text{ even}}}^{2\lfloor j/2\rfloor}(i-2)k_i/2}}\prod_{\substack{i=6\\ i\text{ even}}}^{2\lfloor j/2\rfloor}{2l_i\choose l_i+k_i}\Bigg).\nonumber\eea
}Following the same steps as in Section (\ref{section enumeration on square lattice}) regarding the number $C_{\bf{n}}(A)$   of closed random walks of length ${\bf{n}}=2n$ enclosing on the honeycomb lattice an algebraic area $A$, i.e., considering on the one hand \be\nonumber\sum_A C_{{\bf{n}}}(A) \Q^{A } = {1\over 2q} \text{Tr} H_{2q}^{\bf n},\ee 
which is the anologous of (\ref{Hof Cn and Tr})  for the honeycomb Hamiltonian (\ref{it})  (where the factor $1/q$ is replaced by  $1/(2q)$ in view of a proper normalisation  over the $2q$  states), and  on the other hand  \be \text{Tr} H_{2q}^{\bf n}=2n(-1)^{n+1} b(n),\nonumber \ee which generalizes (\ref{i}), the  expressions  above directly lead  to an  algebraic area enumeration  similar to the  square lattice walks  enumeration (\ref{countingcount}). 

In the sequel, we will consider  $d_q$ in terms
of the original (undiluted) Hofstadter spectrum $s_k$. In that case, the $g=2$ exclusion interpretation does not hold anymore and has to be traded for a mixture of $g=2$ and $g=1$ statistics, as we are going to show in detail.}
 
\subsection{Modified statistics for the spectral function  $s_k$}

If we insist on keeping $s_k$ as the spectral function,
the first few $Z(n)$ rewrite as 
\be Z(1)=\sum _{i=1}^q \big(1+s_{i}\big)\nonumber,\ee
\bea
Z(2)=&+&\sum _{i=1}^{q-1} \sum _{j=1}^i \big(1+s_{i+1}\big) \big(1+s_{j}\big)\nonumber\\&-&\sum _{i=1}^{q-1} s_{i}\nonumber,
\eea
\bea
Z(3)=&+&\sum _{i=1}^{q-2} \sum _{j=1}^i \sum _{k=1}^j \big(1+s_{i+2}\big) \big(1+s_{j+1}\big)
   \big(1+s_{k}\big)\nonumber\\&-&\sum _{i=1}^{q-2} \sum _{j=1}^i \big(1+s_{i+2}\big) s_{j}\nonumber\\&-&\sum _{i=1}^{q-2}
   \sum _{j=1}^i s_{i+1} \big(1+s_{j}\big),\nonumber
\eea
\bea
Z(4)=&+&\sum _{i=1}^{q-3} \sum _{j=1}^i \sum _{k=1}^j \sum _{l=1}^k \big(1+s_{i+3}\big)
   \big(1+s_{j+2}\big) \big(1+s_{k+1}\big) \big(1+s_{l}\big)\nonumber\\&-&\sum _{i=1}^{q-3} \sum _{j=1}^i \sum _{k=1}^j
   \big(1+s_{i+3}\big) \big(1+s_{j+2}\big) s_{k}\nonumber\\&-&\sum _{i=1}^{q-3} \sum _{j=1}^i \sum _{k=1}^j
   \big(1+s_{i+3}\big) s_{j+1} \big(1+s_{k}\big)\nonumber\\&-&\sum _{i=1}^{q-3} \sum _{j=1}^i \sum _{k=1}^j
    s_{i+2} \big(1+s_{j+1}\big) \big(1+s_{k}\big)\nonumber\\&+&\sum _{i=1}^{q-3} \sum _{j=1}^i s_{i+2} s_{j}\nonumber,
\eea
\bea
Z(5)= &+&\sum _{i=1}^{q-4} \sum _{j=1}^i \sum _{k=1}^j \sum _{l=1}^k \sum _{m=1}^l \big(1+s_{i+4}\big)\big(1+s_{j+3}\big)\big(1+s_{k+2}\big)\big(1+s_{l+1}\big)\big(1+s_{m}\big)\nonumber\\
&-&\sum _{i=1}^{q-4} \sum _{j=1}^i \sum _{k=1}^j \sum _{l=1}^k \big(1+s_{i+4}\big)\big(1+s_{j+3}\big)\big(1+s_{k+2}\big)s_{l}\nonumber\\
&-&\sum _{i=1}^{q-4} \sum _{j=1}^i \sum _{k=1}^j \sum _{l=1}^k \big(1+s_{i+4}\big)\big(1+s_{j+3}\big)s_{k+1}\big(1+s_{l}\big)\nonumber\\
&-&\sum _{i=1}^{q-4} \sum _{j=1}^i \sum _{k=1}^j \sum _{l=1}^k \big(1+s_{i+4}\big)s_{j+2}\big(1+s_{k+1}\big)\big(1+s_{l}\big)\nonumber\\
&+&\sum _{i=1}^{q-4} \sum _{j=1}^i \sum _{k=1}^j \big(1+s_{i+4}\big)s_{j+2}s_{k}\nonumber\\
&-&\sum _{i=1}^{q-4} \sum _{j=1}^i \sum _{k=1}^j \sum _{l=1}^k s_{i+3}\big(1+s_{j+2}\big)\big(1+s_{k+1}\big)\big(1+s_{l}\big)\nonumber\\
&+&\sum _{i=1}^{q-4} \sum _{j=1}^i \sum _{k=1}^j s_{i+3}\big(1+s_{j+2}\big)s_{k}\nonumber\\
&+&\sum _{i=1}^{q-4} \sum _{j=1}^i \sum _{k=1}^j s_{i+3}s_{j+1}\big(1+s_{k}\big)\nonumber.
\eea
We infer that in general} the $Z(n)$'s are combinations of nested multiple sums of products of $(1+s_{k})$ and $s_k$ such that
\begin{itemize}
\item
The rightmost factor is either $s_{k}$ or $\big(1+s_{k}\big)$.  
\item
Any factor multiplying $s_i$ immediately on its left obeys $g=2$ exclusions, i.e., 
$\sum_i\sum_j s_{i}s_{j}$ or $\sum_i\sum_j\big(1+s_{i}\big)s_{j}$ where  $i-j\geq 2$.
\item
Any factor multiplying $\big(1+s_i\big)$ immediately on its left obeys $g=1$ exclusions, i.e.,
$\sum_i\sum_js_{i}\big(1+s_{j}\big)$ or $\sum_i\sum_j\big(1+s_{i}\big)\big(1+s_{j}\big)$ where $i-j\geq 1$.
\item
{{The leftmost factor is either $s_{i+n-2}$ or $\big(1+s_{i+n-1}\big)$ with summation range $\sum_{i=1}^{q-(n-1)}$.}}  
\item
The sign is $\pm (-1)^n$  for even/odd number of factors.
\end{itemize}
From these rules  and the very definition  (\ref{00}) we get the $b(n)$'s  in terms of single sums of products of $s_k$ (up to terms involving  $s_q$ which vanish anyway) with a form a bit more complicated than in the Hofstadter case
\begin{align}\nonumber
b(1)&=\sum_{k=1}^{q-1} s_{k} + \sum_{k=1}^{q} s_{k}^0,\\\nonumber
-b(2)&=\frac{1}{2}\sum_{k=1}^{q-1} s_{k}^2 + 2\sum_{k=1}^{q-1} s_{k} + \frac{1}{2}\sum_{k=1}^{q} s_{k}^0,\\\nonumber
b(3)&=\frac{1}{3}\sum_{k=1}^{q-1} s_{k}^3 + 2\sum_{k=1}^{q-1} s_{k}^2 + \sum_{k=1}^{q-2} {s}_{k+1}{s}_{k} + 3\sum_{k=1}^{q-1} s_{k} + \frac{1}{3}\sum_{k=1}^{q} s_{k}^0,\\\nonumber
-b(4)&=\frac{1}{4}\sum_{k=1}^{q-1} s_{k}^4 + 2\sum_{k=1}^{q-1} s_{k}^3 + \sum_{k=1}^{q-2} {s}_{k+1}^2{s}_{k} + \sum_{k=1}^{q-2} {s}_{k+1}{s}_{k}^2\\\nonumber
& + 5\sum_{k=1}^{q-1} s_{k}^2 + 4\sum_{k=1}^{q-2} {s}_{k+1}{s}_{k} + 4\sum_{k=1}^{q-1} s_{k} + \frac{1}{4}\sum_{k=1}^{q} s_{k}^0,\\\nonumber
b(5)&=\frac{1}{5}\sum_{k=1}^{q-1} s_{k}^5 + 2\sum_{k=1}^{q-1} s_{k}^4 + \sum_{k=1}^{q-2} {s}_{k+1}^3{s}_{k} + \sum_{k=1}^{q-2} {s}_{k+1}^2{s}_{k}^2 + \sum_{k=1}^{q-2} {s}_{k+1}{s}_{k}^3\\\nonumber
& + 7\sum_{k=1}^{q-1} s_{k}^3 + 6\sum_{k=1}^{q-2} {s}_{k+1}^2{s}_{k} + 6\sum_{k=1}^{q-2} {s}_{k+1}{s}_{k}^2 + \sum_{k=1}^{q-3} {s}_{k+2}{s}_{k+1}{s}_{k}\\
& + 10\sum_{k=1}^{q-1} s_{k}^2 + 10\sum_{k=1}^{q-2} {s}_{k+1}{s}_{k} + 5\sum_{k=1}^{q-1} s_{k} + \frac{1}{5}\sum_{k=1}^{q} s_{k}^0
\nonumber,\end{align}
i.e., 
\be 
b(n) = (-1)^{n+1} \hskip -0.3cm \sum_{\substack{ l_1, l_2, \ldots, l_j\\ {\rm composition}\;{\rm of}\;  n'= 0, 1, 2, \ldots, n \\ j \leq \min ( n', n-n'+1)}}\hskip -0.4cm 
c_n(l_1,l_2,\ldots,l_{j} )\sum _{k=1}^{q-j} s^{l_{j}}_{k+j-1}\cdots s^{l_2}_{k+1} {s}^{l_1}_k\label{hex bn}\nonumber.
\ee

The new combinatorial coefficients $c_n(l_1,l_2,\ldots,l_{j})$ are labeled by the  compositions
of $n'=0, 1, 2, \ldots, {{n}}$ with a number of parts $j \leq \min (n',n-n'+1)$ (by convention the unique composition of $n'=0$ has only one part and the trigonometric sum becomes $\sum_{k=1}^{q}s^0_{k}$). Since the number of compositions of an integer $n'$ with $j$ parts is  ${n'-1\choose j-1}$, the total number of such   compositions is
\be 1+\sum_{n'=1}^{n}\sum_{j=1}^{\min (n',n-n'+1)}{n'- 1\choose j-1}=1+\sum_{j=1}^{\lfloor (n+1)/2 \rfloor}\sum_{n'=j}^{n-j+1}{n'- 1\choose j-1}=\sum_{j=0}^{\lfloor (n+1)/2 \rfloor}{n-j+ 1\choose j}=F_{n+2}\nonumber.\ee
Note that the Fibonacci  number $F_{n+2}$ is also the number of compositions of $(n+1)$ with only parts $1$ and $2$. We obtain for the $c_n(l_1,l_2,\ldots,l_{j})$'s 
\begin{footnotesize}
\begin{align}\nonumber
c_n(0)&=\frac{1}{n},\\\nonumber
c_n(l_1)&=\frac{1}{l_1}{{n+l_1-1}\choose{2l_1-1}},\\ \nonumber
c_n(l_1,l_2)&=\frac{1}{l_1 l_2}\sum_{m=0}^{\min(l_1,l_2)}m{{l_1}\choose{m}}{{l_2}\choose{m}}{{n+l_1+l_2-m-1}\choose{2(l_1+l_2)-1}},\\
c_n(l_1,l_2,\ldots,l_j)&=\frac{1}{l_1 l_2\ldots l_j}\sum_{m_1=0}^{\min(l_1,l_2)}\sum_{m_2=0}^{\min(l_2,l_3)}\cdots\sum_{m_{j-1}=0}^{\min(l_{j-1},l_{j})}\bigg({\prod_{i=1}^{j-1} m_i}{{l_i}\choose{m_i}}{{l_{i+1}}\choose{m_i}}\bigg){{n+\sum_{i=1}^{j}l_i-\sum_{i=1}^{j-1}m_i-1}\choose{2\sum_{i=1}^{j}l_i-1}}\label{hex coef c}.
\end{align}
\end{footnotesize}
We also note that by ignoring the $n$-dependant  binomial ${{n+\sum_{i=1}^{j}l_i-\sum_{i=1}^{j-1}m_i-1}\choose{2\sum_{i=1}^{j}l_i-1}}$ in the sums (\ref{hex coef c})  one recovers the $c(l_1,l_2,\ldots,l_j)$ in (\ref{Hof coef c}), that is, 
\be c_n(l_1,l_2)\to\frac{1}{l_1 l_2}\sum_{m=0}^{\min(l_1,l_2)}m{{l_1}\choose{m}}{{l_2}\choose{m}} =\frac{{l_1+l_2\choose l_1}}{l_1+l_2}\nonumber,\ee
and thus by factorization
\begin{footnotesize}
\begin{align}\nonumber
c_n(l_1,l_2,\ldots,l_j)\to\frac{1}{l_1 l_2\ldots l_j}\sum_{m_1=0}^{\min(l_1,l_2)}\sum_{m_2=0}^{\min(l_2,l_3)}\cdots\sum_{m_{j-1}=0}^{\min(l_{j-1},l_{j})}{\prod_{i=1}^{j-1} m_i}{{l_i}\choose{m_i}}{{l_{i+1}}\choose{m_i}}=\frac{{l_1+l_2\choose l_1}}{l_1+l_2}\;\; l_2\frac{{l_2+l_3\choose l_2}}{l_2+l_3}\;\cdots \;\; l_{{j}-1}\frac{{l_{{j}-1}+l_{j}\choose l_{{j}-1}}}{l_{{j}-1}+l_{j}}\nonumber.
\end{align}
\end{footnotesize}
We  also have
\be n \sum_{l=0}^{n}{c_n(l)}=F_{2n+1}+F_{2n-1}-1,\nonumber\ee
where again a Fibonacci  counting appears,  and
 \be
n\hskip -0.3cm \sum_{\substack{l_1, l_2, \ldots, l_{j} \\ { \rm composition}\;{\rm of}\;n'\\ j \leq \min ( n', n-n'+1)}} \hskip -0.4cm 
c_n(l_1,l_2,\ldots,l_j)={n \choose n'}^2,
\nonumber\ee
from which we infer
\be n\hskip-0.6cm\sum_{\substack{ l_1, l_2, \ldots, l_j \\ {\rm composition}\;{\rm of}\;  n'= 0, 1, 2, \ldots, n \\  j \leq \min ( n', n-n'+1)}}\hskip -0.5cm c_n(l_1,l_2,\ldots,l_j)={{2n}\choose{n}}.\nonumber\ee
Again using (\ref{counting}), the counting of closed honeycomb lattice walks of length $2n$ is recovered
\bea \nonumber
& & n \hskip -0.6cm\sum_{\substack{ l_1, l_2, \ldots, l_j \\ {\rm composition}\;{\rm of}\; n'= 0, 1, 2, \ldots, n \\  j \leq \min ( n', n-n'+1)}} \hskip -0.5cm c_n(l_1,l_2,\ldots,l_j){2(l_1+l_2+\ldots+l_j)\choose l_1+l_2+\ldots+l_j}\\ \nonumber
&=& \sum_{n'=0}^n\left( n\hskip -0.3cm \sum_{\substack{l_1, l_2, \ldots, l_{j} \\ { \rm composition}\;{\rm of}\;n'\\ j \leq \min ( n', n-n'+1)}} \hskip -0.4cm 
c_n(l_1,l_2,\ldots,l_j){2(l_1+l_2+\ldots+l_j)\choose l_1+l_2+\ldots+l_j}\right)\\
&=& \sum_{n'=0}^{n}{{n}\choose{n'}}^2{{2n'}\choose{n'}}.\nonumber
\eea
\subsection{Algebraic area enumeration on the honeycomb lattice}\label{section enumeration hex}

Remembering that the spectrum of $H_q$ is the square of that of the honeycomb Hamiltonian $H_{2q}$, the generating function for  the number $C_{\bf n}(A)$ of closed  walks of length ${\bf n}=2n$ enclosing an algebraic area $A$  can as well  be  given in terms of the  trace of $H_q^{n}$ weighted  by $1/q$, i.e.,
\be \nonumber
\sum_A C_{\bf n}(A)\Q^A=\frac{1}{q}\text{Tr}H^n_q,
\ee 
%
where now, following again the steps of Section (\ref{section enumeration on square lattice}),
\be\text{Tr} H_q^n = (-1)^{n+1} n b(n)\label{ole}\nonumber.
\ee
{We arrive at the conclusion that on the honeycomb lattice the $C_{\bf n}(A)$'s are

{\small{
\bea
\
C_{\bf n}(A)&=& n \hskip -0.5cm \sum_{\substack{l_1, l_2, \ldots, l_{j} \\ { \rm composition}\;{\rm of}\; n'=0,1,2,\ldots,n\\ j \leq \min (n', n-n'+1)}} \hskip -0.8cm c_n(l_1,l_2,\ldots,l_j)\nonumber\\ && \sum_{k_3=-l_3}^{l_3}\sum_{k_4=-l_4}^{l_4}\cdots\sum_{k_{j}=-l_j}^{l_{j}}{2l_1\choose {l_1+A+\sum_{i=3}^{j}(i-2)k_i}}{2l_2\choose {l_2-A-\sum_{i=3}^{j}(i-1)k_i}}\prod_{i=3}^{j}{2l_i\choose l_i+k_i}
\nonumber\eea}}
with the $c_n(l_1,l_2,\ldots,l_{j})$'s given  in (\ref{hex coef c}) and the algebraic area bounded{\footnote{The sequence OEIS A135711 states that the minimal perimeter of a polyhex with $A$ cells is  $2 \lceil \sqrt{12A-3} \rceil$. The maximum $A$ for walks of length $2n$  is then $\lfloor (n^2+3)/12 \rfloor$.}  by $\lfloor(n^2 + 3)/12\rfloor$. }}

A few examples of {${1\over q}\text{Tr} H^{n}_q$} are listed below, and the corresponding $C_{\bf{n}}(A)$
are listed in Table \ref{hex Cn table}.
\begin{align*}
&\frac{1}{q}\text{Tr}H_q=3,\\
&\frac{1}{q}\text{Tr}H_q^2=15,\\
&\frac{1}{q}\text{Tr}H_q^3=3\left( 29+2\cos\frac{2\pi p}{q}\right),\\
&\frac{1}{q}\text{Tr}H_q^4=3\left(181+32\cos\frac{2\pi p}{q}\right),\\
&\frac{1}{q}\text{Tr}H_q^5=3\left(1181+360\cos\frac{2\pi p}{q}+10\cos\frac{4\pi p}{q}\right),\\
&\frac{1}{q}\text{Tr}H_q^6=3\left(7953+3520\cos\frac{2\pi p}{q}+242\cos\frac{4\pi p}{q}+8\cos\frac{6\pi p}{q}\right),\\
&\frac{1}{q}\text{Tr}H_q^7=3\left(54923+32032\cos\frac{2\pi p}{q}+3710\cos\frac{4\pi p}{q}+266\cos\frac{6\pi p}{q}+14\cos\frac{8\pi p}{q}\right).
\end{align*}
\begin{table}[H]
\begin{center}
\begin{tabular}{|r|c|c|c|c|c|c|c|}
\hline
 & ${\bf n}=2$ & $4$ & $6$ & $8$ & $10$ & $12$ & $14$ \\ \hline
$A=~~0$ & $3$ & $15$ & $87$ & $543$ & $3543$ & $23859$ & $164769$ \\ \hline
$\pm 1$ & & & $6$ & $96$ & $1080$ & $10560$ & $96096$ \\ \hline
$\pm 2$ & & & & & $30$ & $726$ & $11130$ \\ \hline
$\pm 3$ & & & & & & $24$ & $798$ \\ \hline
$\pm 4$ & & & & & & & $42$ \\ \hline
total counting & $3$ & $15$ & $93$ & $639$ & $4653$ & $35169$ & $272835$\\ \hline
\end{tabular}
\caption{\label{hex Cn table} $C_{{\bf{n}}}\left( A\right)$ up to  ${{\bf{n}}}= 14$ for honeycomb lattice walks of length {\bf n}.}
\end{center}
\end{table}
\newpage

\section{Conclusions}

We demonstrated that the area counting of honeycomb walks derives from an exclusion statistics $g=2$ system
with a ``diluted Hofstadter'' spectrum. This fact calls for a more detailed justification: in previous works \cite{Ouvry and Polychronakos 2019, Ouvry and Polychronakos 2020}, two of the
authors had shown that lattice walks that map to exclusion statistics are of the general form
\be \nonumber
H = f(u)\, v + v^{1-g}\, g(u)
\ee
with $u, v$ the quantum torus matrices and $f(u), g(u)$ scalar functions.
The honeycomb Hamiltonian is apparently not of this form. However, the expression
of a walk in terms of a Hamiltonian is not unique: alternative versions corresponding to modular transformations on the
lattice, or, equivalently, alternative realizations of the quantum torus algebra, can exist. We expect that an alternative
realization of the honeycomb Hamiltonian $H_{2q}$ that makes its connection to $g=2$ statistics and the
diluted spectral function $S_k$ manifest does exist, and is related to the form given in Section (\ref{Honeycomb Hamiltonian}) by a unitary transformation.
The identification of this transformation and the alternative form of $H_{2q}$ is an interesting open question.

Further, the anisotropic honeycomb Hamiltonian with general transition amplitudes $a,b,c$, is of physical interest.
The corresponding generating function of lattice walks would depend on these parameters and would ``count''
the number of moves in the three different lattice directions $U,V,W$ separately. The calculation of this generalized
generating function through traces of powers of the Hamiltonian appears to be within reach using the methods
and techniques of this paper and constitutes a subject for further investigation.

\section*{Acknowledgements}
L.G. acknowledges the financial support of China Scholarship Council. A.P. acknowledges the hospitality of LPTMS,
Universit\'e Paris-Saclay, during the early phases of this work.

\newpage
\appendix
\section*{Appendices}
\addcontentsline{toc}{section}{Appendices}
\renewcommand{\thesubsection}{\Alph{subsection}}

\subsection{$Z(n)$ for square lattice walks}

We denote $Z(n)$ as $Z_q(n)$ to include its dependence on $q$.

Substituting $\displaystyle{d_q=\sum_{n=0}^{\lfloor q/2 \rfloor}(-1)^n Z_{q}(n) z^{2n}}$ into \eqref{Hofrec} and equating the coefficient of $z^{2n}$ on both sides, we get
\bea \nonumber
Z_{q}(n) &=& Z_{q-1}(n)+s_{q-1}Z_{q-2}(n-1) \\ \nonumber
&=& Z_{q-2}(n)+s_{q-2}Z_{q-3}(n-1)+s_{q-1}Z_{q-2}(n-1)\\ \nonumber
&=& \cdots \\ \nonumber
&=& Z_{1}(n)+\sum_{m=0}^{q-2}s_{m+1}Z_{m}(n-1).\eea
Since $Z_{m}(n-1)=0$ for $n-1 > \lfloor m/2 \rfloor$, i.e., $m<2n-2$,
we obtain
\be Z_q(n)=\sum_{m=2n-2}^{q-2}s_{m+1}Z_{m}(n-1) {\label{Appendix Hof Zn recursion}}\nonumber\ee
with $Z_{q}(0)=1$.

Thus,
\bea \nonumber
Z_{q}(1)&=&\sum_{m=0}^{q-2}s_{m+1}Z_m(0) \\ \nonumber
&=&\sum_{k_1=1}^{q-1}s_{k_1},
\eea
\bea \nonumber Z_{q}(2)&=& \sum_{m=2}^{q-2}s_{m+1}Z_m(1) \\ \nonumber
&=&\sum_{m=2}^{q-2}\sum_{k_1=1}^{m-1}s_{m+1}s_{k_1} \\ \nonumber
&=& \sum_{k_1=1}^{q-3}\sum_{k_2=1}^{k_1}s_{k_1+2}s_{k_2},
\eea
\bea \nonumber Z_{q}(3)&=&\sum_{m=4}^{q-2}s_{m+1}Z_m(2) \\ \nonumber
&=&\sum_{m=4}^{q-2}\sum_{k_1=1}^{m-3}\sum_{k_2=1}^{k_1}s_{m+1}s_{k_1+2}s_{k_2}\\ \nonumber
&=&\sum_{k_1=1}^{q-5}\sum_{k_2=1}^{k_1}\sum_{k_3=1}^{k_2}s_{k_1+4}s_{k_2+2}s_{k_3}.
\eea

\noindent
The formula
(\ref{Hof Zn bis}) 
can be then proven by mathematical induction, where we check
\bea \nonumber
Z_{q}(n+1) &=& \sum_{m=2n}^{q-2}s_{m+1}Z_{m}(n)\\ \nonumber
&=& \sum_{m=2n}^{q-2}\sum_{k_1=1}^{m-2n+1}\sum_{k_2=1}^{k_1}\sum_{k_3=1}^{k_2}\cdots\sum_{k_{n}=1}^{k_{n-1}}s_{m+1}s_{k_1+2n-2}\cdots s_{k_{n-1}+2}s_{k_{n}} \\ \nonumber
&=& \sum_{k_1=1}^{q-2n-1}\sum_{k_2=1}^{k_1}\sum_{k_3=1}^{k_2}\sum_{k_4=1}^{k_3}\cdots\sum_{k_{n+1}=1}^{k_{n}}s_{k_1+2n}s_{k_2+2n-2}\cdots s_{k_{n}+2}s_{k_{n+1}}.
\eea

\subsection{Examples of algebraic area enumeration of random walks on the square lattice}
A few examples of ${1\over q}\text{Tr} H^{\bf{n}}_q$ and the corresponding    $C_{\bf n}(A)$'s are listed below and  in Table \ref{Hof Cn table}. 
{\footnotesize \begin{align*}
&\frac{1}{q}\text{Tr}H_q^2=4,\\
&\frac{1}{q}\text{Tr}H_q^4=4\left(7+2\cos\frac{2\pi p}{q}\right),\\
&\frac{1}{q}\text{Tr}H_q^6=4\left(58+36\cos\frac{2\pi p}{q}+6\cos\frac{4\pi p}{q}\right),\\
&\frac{1}{q}\text{Tr}H_q^8=4\left(539+504\cos\frac{2\pi p}{q}+154\cos\frac{4\pi p}{q}+24\cos\frac{6\pi p}{q}+4\cos\frac{8\pi p}{q}\right),\\
&\frac{1}{q}\text{Tr}H_q^{10}=4\left(5486+6580\cos\frac{2\pi p}{q}+2770\cos\frac{4\pi p}{q}+780\cos\frac{6\pi p}{q}+210\cos\frac{8\pi p}{q}+40\cos\frac{10\pi p}{q}+10\cos\frac{12\pi p}{q}\right).
\end{align*}}
\begin{table}[H]
\begin{center}
\begin{tabular}{|r|c|c|c|c|c|}
\hline
 & ${\bf n}=2$ & $4$ & $6$ & $8$ & $10$ \\ \hline
$A=~~0$ & $4$ & $28$ & $232$ & $2156$ & $21944$ \\ \hline
$\pm 1$ &  & $8$ & $144$ & $2016$ & $26320$ \\ \hline
$\pm 2$ &  &  & $24$ & $616$ & $11080$ \\ \hline
$\pm 3$ &  &  &  & $96$ & $3120$ \\ \hline
$\pm 4$ &  &  &  & $16$  & $840$\\ \hline
$\pm 5$ &  &  &  &  & $160$ \\ \hline
$\pm 6$ &  &  &  &  & $40$ \\ \hline
counting & $4$ & $36$ & $400$ & $4900$ & $63504$ \\ \hline
\end{tabular}
\caption{\label{Hof Cn table} $C_{{\bf{n}}}\left( A\right)$ up to  ${{\bf{n}}}= 10$ for square lattice walks of length $\bf n$.}
\end{center}
\end{table}

\subsection{Representation of the honeycomb algebra}\label{Appendix UVW proof}

Define three new operators $u,v,\sigma$ as
\bea\nonumber
&&\sigma=\Q^{-1/2}UVW,~ u=U\sigma,~ v=V\sigma \\
\Rightarrow ~~~&&U=u\sigma,~ V=v\sigma,~ W=\Q^{1/2}v\sigma u \nonumber.
\eea
From the honeycomb algebra (\ref{honey}) we see
that $\sigma$, $u$ and $v$ are all unitary and satisfy
\be
vu=\Q uv,~ u\sigma=\sigma u^{-1},~ \sigma v^{-1}=v\sigma,~ \sigma^2 = 1 
\label{newhoney}.\ee
Since $U$, $V$ and $W$ can be uniquely expressed in terms of $\sigma$, $u$ and $v$, it is sufficient to derive the irreducible representation (``irrep" for short) of $u$, $v$ and $\sigma$.

Operators $u$ and $v$ satisfy the quantum torus algebra and have a
$q$-dimensional irrep if $\Q=\exp(2{\rmi}\pi p/q)$.
However, $\sigma$ can be embedded within this irrep only for specific
values of the Casimirs $u^q = {\rme}^{{\rmi}\phi}$ and $v^q = {\rme}^{{\rmi}\theta}$. Indeed, assuming $\sigma$ acts within this irrep,
\be\nonumber
u^q = \sigma u^q \sigma = u^{-q} ~~~ \Rightarrow ~~~ {\rme}^{{\rmi}\phi} = {\rme}^{-{\rmi}\phi}.
\ee
So $\phi$ can only be $0$ or $\pi$ (mod $2\pi$), and similarly for $\theta$. For  $\theta,\phi \in \{0,\pi\}$ we can show
that the irrep of (\ref{newhoney}) is unique up to unitary transformations, and up to the algebra automorphism
$\sigma \to -\sigma$, and is given by the action on basis states $\ket{n}$
\bea
u\ket{n}=&{\rme}^{{\rmi}({\phi}+{2\pi p}n)/q}\ket{n} &,~~ n=0,1,\ldots,q-1, \nonumber\\
v\ket{n}=&{\rme}^{{\rmi}{\theta}/{q}}\ket{n\bb -\bb 1}&,~~ \ket{-1}\equiv\ket{q-1},\nonumber\\
\sigma \ket{n}=& {\rme}^{{\rmi}\theta(2n-r)/q} \ket{r\bb -\bb n} &,~~rp+{\phi/\pi} = 0~ ({\text{mod}}~q) \nonumber
\label{qdim}.\eea
The ``pivot'' $r$ in the inversion action of $\sigma$ is $r=0$, if $\phi=0$, and the primary solution of the Diophantine equation
$kq-rp=1$, if $\phi=\pi$. The momenta $q k_x =  \theta$ and $q k_y = \phi$ in this irrep are quantized as
\be
k_x = {\pi n_x\over q},~~k_y = {\pi n_y\over q},~~~ n_x , n_y \in   \mathbb{Z}.
\ee

For either $\theta$ or $\phi \notin \{0,\pi\}$ the irrep of (\ref{newhoney}) must decompose into more than one $q$-dimensional
irreps of the quantum torus algebra $u,v$ with $\sigma$ mixing the irreps. The minimal irrep of the full algebra (\ref{newhoney})
involves 2 irreps of the torus algebra, all other situations being reducible. Representing all operators in block diagonal form
in the space of the two irreps $u_i , v_i$, $i=1,2$, with Casimirs $u_i^q = {\rme}^{{\rmi} \phi_i}$, $v_i^q = {\rme}^{{\rmi} \theta_i}$,
\be
u=\begin{pmatrix}
u_1 & 0 \\
0 & u_2\\
\end{pmatrix},~~
v=\begin{pmatrix}
v_1 & 0 \\
0 & v_2\\
\end{pmatrix},~~
\sigma=\begin{pmatrix}
A~ & B \\
B^\dagger & C\\
\end{pmatrix},
\nonumber\ee
and implementing the relations $\sigma u^q \sigma = u^{-q}$, $\sigma v^q \sigma = v^{-q}$ leads to
\bea
\left({\rme}^{{\rmi}\phi_1} - {\rme}^{-{\rmi}\phi_1} \right) A &=  \left({\rme}^{{\rmi}\phi_2} - {\rme}^{-{\rmi}\phi_2} \right) C = \left({\rme}^{{\rmi}\phi_1} - {\rme}^{-{\rmi}\phi_2} \right) B &= 0,\nonumber\\
\left({\rme}^{{\rmi}\theta_1} - {\rme}^{-{\rmi}\theta_1} \right) A &=  \left({\rme}^{{\rmi}\theta_2} - {\rme}^{-{\rmi}\theta_2} \right) C \hskip 0.05cm 
= \left({\rme}^{{\rmi}\theta_1} - {\rme}^{-{\rmi}\theta_2}\hskip 0.05cm \right) \hskip 0.03cm B &= 0.
\nonumber\eea
Since not both of $\phi_1,\phi_2$ and of $\theta_1,\theta_2$ can be $0$ or $\pi$, the above relations imply $A=C=0$.
$\sigma^2=1$ then implies $B^\dagger B = 1$, and the last equalities above require $\phi_1 = - \phi_2$, $\theta_1 = -\theta_2$.
Further, a unitary transformation
\be\nonumber
S = \begin{pmatrix}
\bb B^\dagger\, & 0 \\
0~\,  & 1\\
\end{pmatrix},~~ u \to S u S^{-1},~ v \to S v S^{-1},~ \sigma \to S \sigma S^{-1}
\ee
eliminates $B$ in $\sigma$, and $\sigma u\sigma = u^{-1} ,~ \sigma v\sigma = v^{-1}$ imply $u_1 = u_2^{-1}$, $v_1 = v_2^{-1}$. Altogether, the irrep of (\ref{newhoney}) for two arbitrary Casimirs $\phi = \phi_1=-\phi_2$, 
$\theta = \theta_1=-\theta_2$, is given by the $2q$-dimensional matrices
\be 
u=\begin{pmatrix}
u_o & 0\bb \\
0 & ~ u_o^{-1}\bb\\
\end{pmatrix}
,~~
v=\begin{pmatrix}
v_o & 0\bb \\
0 & ~v_o^{-1}\bb\\
\end{pmatrix}
,~~
\sigma=\begin{pmatrix}
0 & ~1\\
1 & ~0\\
\end{pmatrix},
\label{2qdim}\ee
where $u_o$ and $v_o$ are the basic $q$-dimensional quantum torus irrep with Casimirs ${\rme}^{{\rmi}\phi}$ and ${\rme}^{{\rmi}\theta}$.
Finally, from (\ref{newhoney}) we obtain the corresponding irreducible forms for $U,V,W$
\be\nonumber
U=\begin{pmatrix}
0~ & u_o \\
u_o^{-1} & 0\\
\end{pmatrix}
,~~
V=\begin{pmatrix}
0~ & v_o \\
v_o^{-1} & 0\\
\end{pmatrix}
,~~
W=\Q^{1/2} \begin{pmatrix}
0~ &~ v_o u_o^{-1}\\
 v_o^{-1} u_o& 0\\
\end{pmatrix}.
\ee

We conclude with a demonstration that the above representation becomes reducible if $\phi,\theta \in \{0,\pi\}$. In that
case, as we demonstrated before in (\ref{qdim}), there is a $q\times q$ matrix $\sigma_o$ (to be distinguished from the
$2q \times 2q$ matrix $\sigma$ in (\ref{2qdim}) above) satisfying (\ref{newhoney}) for the matrices $u_o$ and $v_o$.
Performing the unitary transformation
\be\nonumber
S_o = {1\over \sqrt{2}} \begin{pmatrix}
1 & -\sigma_o \\
\sigma_o & 1\\
\end{pmatrix}
\ee
on all matrices, and using $\sigma_o u_o \sigma_o = u_o^{-1}$ etc., we obtain
\be\nonumber
u=\begin{pmatrix}
u_o & 0 \\
0 & ~u_o^{-1}\\
\end{pmatrix}
,~~
v=\begin{pmatrix}
v_o & 0 \\
0 & ~v_o^{-1}\\
\end{pmatrix}
,~~
\sigma=\begin{pmatrix}
\sigma_o & ~0\\
0 & -\sigma_o\\
\end{pmatrix},
\ee
or
\be\nonumber
U=\begin{pmatrix}
u_o \sigma_o & 0 \\
0 & -\sigma_o u_o\\
\end{pmatrix}
,~~
V=\begin{pmatrix}
v_o \sigma_o & 0 \\
0 & -\sigma_o v_o\\
\end{pmatrix}
,~~
W= \Q^{1/2} \begin{pmatrix}
v_o u_o^{-1} \sigma_o & 0\\
0 & -\sigma_o v_o u_o^{-1}\\
\end{pmatrix}
\ee
reducing to the direct sum of two $q$-dimensional irreps.

\subsection{$Z(n)$ for honeycomb lattice walks}
We denote $Z(n)$ as $Z_q(n)$ to include its dependence on $q$. 

Substituting $\displaystyle{d_q=\sum_{n=0}^{q} (-1)^n Z_{q}(n)z^{2n}}$ into \eqref{hexrec} and equating the coefficient of $z^{2n}$ on both sides, we get
\begin{small}
\bea \nonumber
Z_{q}(n)&=&Z_{q-1}(n)+\big(1+s_q\big)Z_{q-1}(n-1)-s_{q-1}Z_{q-2}(n-2) \\ \nonumber
&=& Z_{q-2}(n)+\big(1+s_{q-1}\big)Z_{q-2}(n-1)+\big(1+s_q\big)Z_{q-1}(n-1)-s_{q-2}Z_{q-3}(n-2)-s_{q-1}Z_{q-2}(n-2)\\ \nonumber
&=& \cdots \\ \nonumber
&=& Z_{1}(n)+\sum_{m=1}^{q-1}\big(1+s_{m+1}\big)Z_{m}(n-1)-\sum_{m=0}^{q-2}s_{m+1}Z_{m}(n-2).
\eea
\end{small}Since $Z_{m}(n)=0$ for $n > m$,
we obtain
\be\nonumber Z_q(n)=\sum_{m=n-1}^{q-1}\big(1+s_{m+1}\big)Z_{m}(n-1)-\sum_{m=n-2}^{q-2}s_{m+1}Z_{m}(n-2) {\label{Appendix hex Zn recursion}}\ee
with $Z_{q}(0)=1$ and $Z_{q}(j)=0$ for $j<0$.

Thus,
\begin{small}
\bea \nonumber
Z_q(1)&=& \sum_{m=0}^{q-1}\big( 1+s_{m+1}\big) Z_{m}(0)\\ \nonumber
&=& \sum _{k_1=1}^q \big(1+s_{k_1}\big),
\eea
\bea \nonumber
Z_q(2)
&=&\sum_{m=1}^{q-1}\big(1+s_{m+1}\big)Z_{m}(1)-\sum_{m=0}^{q-2}s_{m+1}Z_{m}(0)\\ \nonumber
&=&\sum_{m=1}^{q-1}\sum_{k_1=1}^{m}\big(1+s_{m+1}\big)\big(1+s_{k_1}\big)-\sum_{m=0}^{q-2}s_{m+1}\\ \nonumber
&=&\sum _{k_1=1}^{q-1} \sum _{k_2=1}^{k_1} \big(1+s_{k_1+1}\big) \big(1+s_{k_2}\big)-\sum _{k_1=1}^{q-1} s_{k_1},\eea
\bea \nonumber
Z_q(3)
&=& \sum_{m=2}^{q-1}\big(1+s_{m+1}\big)Z_{m}(2)-\sum_{m=1}^{q-2}s_{m+1}Z_{m}(1)\\ \nonumber
&=& \sum_{m=2}^{q-1}\sum _{k_1=1}^{m-1} \sum _{k_2=1}^{k_1} \big(1+s_{m+1}\big)\big(1+s_{k_1+1}\big) \big(1+s_{k_2}\big)-\sum_{m=2}^{q-1}\sum_{k_1=1}^{{m-1}} \big( 1+s_{m+1} \big)s_{k_1}-\sum_{m=1}^{q-2}\sum_{k_1=1}^{m}s_{m+1}\big( 1+s_{k_1}\big)\\ \nonumber
&=&\sum _{k_1=1}^{q-2} \sum_{k_2=1}^{k_1} \sum _{k_3=1}^{k_2} \big(1+s_{k_1+2}\big) \big(1+s_{k_2+1}\big) \big(1+s_{k_3}\big)-\sum _{k_1=1}^{q-2} \sum_{k_2=1}^{k_1} \big(1+s_{k_1+2}\big) s_{k_2}-\sum _{k_1=1}^{q-2} \sum_{k_2=1}^{k_1} s_{k_1+1} \big(1+s_{k_2}\big).\eea
\end{small}

\end{document}